# Optical control of magnetism in NiFe/VO$_2$ heterostructures

*Guodong Wei,*[1,‡] *Xiaoyang Lin,*[1,2,‡,]* *Zhizhong Si,*[1] *Dong Wang,*[3] *Chen Cai,*[1] *Xinhe Wang,*[1,4] *Xiaofei Fan,*[1] *Yuan Cao,*[1] *Kai Liu,*[5] *Kaili Jiang,*[4] *Zhaohao Wang,*[1] *Na Lei,*[1] *Yanxue Chen,*[3] *Stephane Mangin,*[6] *Weisheng Zhao*[1,2,]*

[1] Fert Beijing Research Institute, School of Microelectronics & Beijing Advanced Innovation Center for Big Data and Brain Computing (BDBC), Beihang University, Beijing 100191, China

[2] Beihang-Goertek Joint Microelectronics Institute, Qingdao Research Institute, Beihang University, Qingdao 266000, China

[3] School of Physics and State Key Laboratory of Crystal Materials, Shandong University, Jinan 250100, China

[4] State Key Laboratory of Low-Dimensional Quantum Physics, Department of Physics & Tsinghua-Foxconn Nanotechnology Research Center, Collaborative Innovation Center of Quantum Matter, Tsinghua University, Beijing 100084, China

[5] School of Materials Science & Engineering, Tsinghua University, Beijing 100084, China

[6] Institut Jean Lamour, UMR 7198, CNRS-Universite de Lorraine, F-54000 Nancy, France

‡These authors contributed equally.

*E-mail: XYLin@buaa.edu.cn (X.Y.L), weisheng.zhao@buaa.edu.cn (W.S.Z)

# Abstract

Optical methods for magnetism manipulation have been considered as a promising strategy for ultralow-power and ultrahigh-speed spin switches, which becomes a hot spot in the field of spintronics. However, a widely applicable and efficient method to combine optical operation with magnetic modulation is still highly desired. Here, the strongly correlated electron material $VO_2$ is introduced to realize phase-transition based optical control of the magnetism in NiFe. The NiFe/$VO_2$ bilayer heterostructure features appreciable modulations in electrical conductivity (55%), coercivity (60%), and magnetic anisotropy (33.5%). Further analyses indicate that interfacial strain coupling plays a crucial role in this modulation. Utilizing this optically controlled magnetism modulation feature, programmable Boolean logic gates (AND, OR, NAND, NOR, XOR, NXOR and NOT) for high-speed and low-power data processing are demonstrated based on this engineered heterostructure. As a demonstration of phase-transition spintronics, this work may pave the way for next-generation electronics in the post-Moore era.

With integrated circuit production approaching its physical limits, spintronics becomes one of the most promising technologies to for information storage and processing in the post-Moore era, owing to its superior properties like nonvolatility, high-speed and low-power consumption[1,2]. The modulation of magnetism, especially multifield control of magnetism like voltage-controlled magnetic anisotropy (VCMA)[3,4], all optical switching (AOS)[5,6], multiferroic[7,8] and ionic[9–11] modulation, has already demonstrated great potential in device performance improvement and application revolution[12]. Meanwhile, optical devices with the advances of high-speed, wide-bandwidth and low-loss have greatly promoted the modern communication industry[13]. Combining the advantages of spintronic and optoelectronic devices may highly improve the energy conversion efficiency, increase the device speed and diversify the functionalities[14]. In this sense, a strategy to realize optical control of magnetic characteristics can certainly promote the emerging device applications.

Strong correlations between electrons enable extraordinary control of physical properties (e.g., magnetism and conductivity) via different strategies (e.g., electric field, magnetic field and stress), which implies numerous possibilities of strongly correlated electron material in various research fields, including high-temperature superconductors, two-dimensional electron gas systems and next-generation spintronics[15]. As a representation of the strongly correlated electron material, $VO_2$ exhibits a complex property change as it transforms from a monoclinic (M1) insulator into a rutile (R) metal at a critical temperature around 340 K (Fig. 1a)[16,17]. The phase transition can also be induced by kinds of ways like electrical gating[12], optothermal effect (Supplementary Fig. S1) or even the photonic effect[19], which has already enriched its device applications. The sub-ps phase transition of $VO_2$ further endows the $VO_2$-related devices with the feature of ultra-fast speed[20,21]. Therefore, a phase-transition based spintronic device will reform the magnetism modulation strategy and give birth to emerging optical-spintronic device applications.

In this work, we study the optical control of magnetism modulation (OCM modulation) in $NiFe/VO_2$ heterostructures that combine a strongly correlated electron material with a magnetic material. Appreciable and reversible modulations in electrical conductivity (55%), coercivity (60%) and magnetic anisotropy (33.5%) have been achieved in these heterostructures during the phase transition of $VO_2$. Further analyses indicate that interfacial strain coupling plays a crucial role in this modulation. Based on these modulation features, the heterostructures are further used to demonstrate multiresistance states and implement spin logic devices with multifield operation capability. As an optical magnetism modulation strategy, the collision and blending of strongly correlated electron materials and spintronic materials may pave the way for next-generation electronics.

# RESULTS

## Heterostructure preparation and magnetism modulation induced by $VO_2$ phase-transition

The $NiFe/VO_2$ bilayer heterostructure used in this work was prepared by two steps. $VO_2$ was first epitaxially deposited on $TiO_2$ substrates by pulsed laser deposition (PLD) to ensure an optimal phase transition. NiFe was then deposited at room temperature by magnetron sputtering to get a tight interface contact. Fig. 1b shows the XRD results of the heterostructure samples. The (200) and (002) peaks of $VO_2$ were detected on the $TiO_2$ (100) and (001) substrates, respectively. The spectrum shows no distinct characteristic peaks for NiFe, which indicates that the NiFe layer is amorphous.

To investigate the magnetic characteristics modulated by the phase transition, hysteresis curves of a NiFe (5 nm)/$VO_2$ (20 nm) sample grown on a $TiO_2$ (100) substrate were measured for the case in which the $VO_2$ is in the monoclinic state (MS, 300 K) and rutile state (RS, 360 K) using a SQUID magnetometer. As illustrated in Fig. 1c, obvious anisotropy can be observed between the results with an in-plane and out-of-plane magnetic field applied. After the $VO_2$ phase transition, the saturation magnetization of the heterostructure decreases from 898 emu/cm$^3$ to 840 emu/cm$^3$. The insets show scanning results obtained over a small region, and shrinking of the coercive and saturation fields can be detected. Remarkably, the decrease is more obvious out-of-plane, enhancing the squareness of the curve around the zero field. The coercive field decreases from 72 Oe to 45 Oe, i.e., a change of 60%. Excluding the effect of a $VO_2$ single layer[22], (Supplementary Fig. S1c), this phase-transition-induced magnetism modulation, which is different from a mere thermal effect (Supplementary Fig. S2c for samples without the $VO_2$ layer), shows great potential for spintronic application[23].

To further explore this modulation effect, the magnetic parameters of the heterostructure before and after the phase transition are processed and summarized in Table 1. We analyzed the in-plane and out-of-plane hysteresis curves of the film and obtained the uniaxial magnetic anisotropy, $K_u$, which increases from -3.55×10$^6$ erg/cm$^3$ to -2.66×10$^6$ erg/cm$^3$. To separate the bulk and interfacial contributions of the anisotropy, the relationship between $K_u$ and the NiFe thickness, $t$, can be written as $K_u=K_b-2\pi M_S^2+K_i/t$ [24], where $K_b$ is the bulk crystalline anisotropy, $K_i$ is the interfacial anisotropy, and $-2\pi M_s^2$ is the demagnetizing field of a uniform sheet of NiFe. In this convention, $K_u < 0$ stands for an in-plane easy axis. Therefore, the increasing tendency indicates an enhancement of the perpendicular magnetic anisotropy after the phase transition[25]. The bulk crystalline anisotropy is negligible because of the amorphous feature of the NiFe film. The interfacial magnetic anisotropy can then be calculated as $K_i=(K_u+2\pi M_s^2)t$, which increases from 0.755 erg/cm$^2$ to 0.885 erg/cm$^2$. These results indicate the possibility of realizing an interfacial perpendicular magnetic anisotropy enhancement in similar heterostructures, which is highly desirable for high-density, nonvolatile magnetic storage[24].

## Optical control of the magnetic and transport properties in the heterostructure

After showing the magnetism modulation effect achieved though phase transition of VO$_2$, we now focus on realizing the optically controlled magnetic modulation by optothermal effect. To achieve this goal, a red laser beam is used as a light source to trigger the phase transition. Fig.2a gives the hysteresis loops of the sample at initial, light-on and light-off states measured by in situ longitudinal magneto-optic Kerr effect (MOKE). Obvious modulation effect can be detected between initial and light-on states which verifies the optically controlled magnetic modulation feature. The light-off state is fully coinciding with the initial state, reflecting that this modulation is totally reversible.

To further study the modulation phenomenon, a phase-transition anisotropic magnetoresistance device (PTAMR device) was fabricated. A schematic illustration of the experimental setup is shown in Fig. 2b. The sample was patterned into a hall bar. Magnetic and transport characteristics of this device were studied via magnetoresistance measurements. Fig. 2c presents the dependence of the device resistance on the light illumination power. Interestingly, the resistance of the heterostructure decreases after an increasement at small illumination power when a small current (1 μA & 10 μA) is applied. For a large measuring current (100 μA), the resistance first rises, then decreases, and then increases again (inset of Fig. 2c). The change rates of the resistance decrease from 55% to almost 0% as the measuring current is increased. In contrast to the simple rising feature under heating in ordinary metals, this complex resistance variation can be explained as the competition between the resistance-temperature dependence of the magnetic metal and the current shunting effect induced by the $VO_2$ phase transition. Normally, illumination and Joule heating will increase the resistance of the device; however, the phase transition of $VO_2$ from insulator to metal reduces the device resistance. As a result, the resistance will first increase and then decrease when the phase transition occurs. In this sense, a small current can highlight the phase-transition feature. Thus, a measuring current of 1 μA is employed to perform the magnetoresistance measurement.

Magnetoresistance curves of the device before and after the $VO_2$ phase transition are shown in Fig. 2d, 2e & 2f. One obvious variation in this process is the shift in the

"valley bottom" (i.e., from approximately 50 Oe to 10 Oe at 1 W/cm$^2$, with a further decrease for 1.6 W/cm$^2$), which agrees with the modulation of the coercivity and magnetic anisotropy in the heterostructure (Fig. 1 & TABLE 1). Moreover, even though the resistance decreases with increasing light power (10% upon the application of 1 W/cm$^2$ and 27% upon the application of 1.6 W/cm$^2$), the device remains sensitive to the magnetic field, i.e., it simultaneously responds to light and magnetic stimuli. These appreciable effects in the heterostructure may enable emerging device applications, which will be discussed in the last section of this paper.

**Strain analysis and mechanism explanation**

To clarify the relationship between the magnetism modulation and the phase transition of VO$_2$, a series of experiments was carried out by MOKE measurements. As illustrated in Fig. 3a, even though the change rates are different, obvious coercive field shrinking can be observed for both heterostructure samples. However, only faint deviations of $H_C$ can be detected for the pristine NiFe sample (Supplementary Fig. S2a). These results suggest that the modulation effect is dominated by the phase transition of VO$_2$, rather than a pure thermal effect of the NiFe layer. We then measured the magnetization-temperature (*M-T*) dependence of the heterostructure sample. As illustrated in Fig. 3b, an abnormal peak can be detected in the M-T curve near the phase-transition critical temperature of VO$_2$. The magnetization then increases with the temperature increase due to the coercive field shrinking. This phenomenon was not detected in the pristine NiFe or VO$_2$ samples (Supplementary Fig. S1c), which further confirms that the magnetism variation is indeed related to the phase transition of the

VO$_2$ layer.

Further experimental evidence indicates that this modulation is an interfacial effect. For heterostructure samples deposited on various TiO$_2$ substrates (Fig. 3a), the experimental data show that the change in $H_C$ reaches 45% for TiO$_2$ (100) but only 25% for TiO$_2$ (001) between the MS and RS. This feature verifies that the modulation depends on the crystalline structure of the VO$_2$ layer. Fig. 3c shows the analysis of the crystal axis conversion of VO$_2$ during its phase transition[19]. Remarkably, a more appreciable change can be achieved on the VO$_2$ layer grown on the TiO$_2$ (100) substrate. On the TiO$_2$ (001) substrate, the in-plane lattice constant changes are approximately 0.4% and 0.2% along the $a_R$ and $b_R$ directions, respectively (Fig. 3d). On the TiO$_2$ (100) substrate, the lattice constant is compressed by approximately 0.8% along $c_R$, and the change rate between $b_R$ and $c_M$ reaches 18.5% with the additional shear force caused by the angle decrease of $\beta$ from 122.6° to 90° (Fig. 3e). As a result, the interfacial strain is enhanced on the VO$_2$/TiO$_2$ (100) sample. This analysis suggests that the different magnetism modulation mechanisms could be attributed to the different magneto-elastic effect induced by the structure change, i.e., the interfacial strain coupling of the heterostructure between VO$_2$ and NiFe. To confirm this conjecture, heterostructures with different NiFe thicknesses were examined (Supplementary Fig. S2b & S2c). The results show that the change in $H_C$ reaches 67% as the thickness decreases to 3 nm and quickly drops to nearly 0% as the thickness increases to 10 nm. This result agrees with reports that the magneto-elastic coupling coefficient of NiFe is anomalously large for thicknesses below 5 nm[26]. Further engineering of this interfacial strain coupling effect in a NiFe/VO$_2$ heterostructure could enable optimized modulation performance for various device applications.

Based on the understanding that the magnetism modulation is mainly due to the interfacial strain, we focused on the origin of the magnetism modulation caused by interfacial strain during the phase transition[27]. Recent research has shown that by engineering the surface, perpendicular or other directions of magnetic anisotropy can be achieved in ultrathin magnetic films[28]. In our phase-transition heterostructure, the lattice variation in the specific plane is quite large, which could lead to some periodic bending and spin reorientation of the NiFe film in the interface region (Supplementary Fig. S3a)[29]. Assuming that the main modulation mechanism is domain wall motion, the coercivity $H_c$ should depend on the domain wall energy[30]:

$$H_c \approx (\partial \gamma / \partial x)_{max} / (2\mu_0 M_s),$$

where $x$ is the position of the domain wall, $M_s$ is the saturation magnetization, $\mu_0$ is the permeability in vacuum and $\gamma$ is the domain wall energy. $\gamma$ can be further written as[31]:

$$\gamma = 4\sqrt{A(K_1 + \lambda_s \sigma)},$$

where $A$ is the exchange constant, $K_1$ is the crystal anisotropy, $\lambda_s$ is the magnetostriction coefficient and $\sigma$ is the applied strain to the sample. Thus, $\gamma$ depends on $\sigma$ and further influences $H_c$.

As shown in Fig. 4a & 4b, the cross-section transmission electron microscopy (cross-section TEM) results of the heterostructure on a TiO$_2$ (100) substrate verify the existence of thin-film bending in the heterostructure and the absence of obvious intermixing at the interface. Some distortion of the crystal orientation in the VO$_2$ layer can be observed near the bending area (Fig. 4a, TEM image; Fig. 4d, nanobeam electron diffraction result), which may be caused by the $VO_2$ phase-transition cooling from a high growth temperature after the PLD deposition. These structure features may further enhance the interface roughness after the phase transition, which may induce reorientation of the spin from in-plane to out-of-plane[26] (Fig. 4e). These results are in agreement with the magnetic characteristic results, i.e., the increased $K_u$ and enhanced squareness of the out-of-plane hysteresis loop. The saturation magnetization decrease and coercive force shrinkage are probably caused by the varied distances between individual magnetic grains caused by the bending interface structure. The intergranular interaction might thus be weakened, which inhibits long-range exchange coupling.

**Multiresistance realization and logic function implementation**

These appreciable optically controlled modulation effects in heterostructures may further enable emerging device applications with multifield modulation capability[32–34]. Fig. 5a & 5b show that the magnetoresistance curves of the heterostructure move to a lower resistance state once illumination is applied. Utilizing this feature, the PTAMR device provides an opportunity to achieve multiresistance states controlled by the magnetic field and light illumination. Fig. 5c demonstrates an example of six different resistance states achieved via synergistic control of the light illumination and magnetic field. Both field and illumination controls enable reversible, repeatable, stable modulation of the device resistance. Further implementation of this phase-transition spintronic effect into other devices, such as giant magnetoresistance (GMR) or tunnel magnetoresistance (TMR), would realize similar but more appreciable performances.

With great potential in information processing, new strategies for spin logic applications are highly desired[35,36]. In the following, we will present a demonstration of multifield programmable logic gates based on the PTAMR device. Fig. 5d shows four resistance states selected from Fig. 5c, corresponding to a light illumination of 1.0 W/cm$^2$. The programmable logic gates include two kinds of input signals, light illumination and magnetic field, that carry the input logic information. The light inputs with illumination on (1.0 W/cm$^2$) and off (0 W/cm$^2$) are defined as input 0 and input 1, respectively. The field input with a magnetic field (400 Oe) and no field (0 Oe) is set as input 1 and input 0, respectively. The computation is read out by measuring the device resistance.

As illustrated in the truth table in Fig. 5d, four different resistance states can be achieved with different inputs. In this case, we can set different threshold resistances to define in the range in which the resistance can be read as logic 1 and that in which the resistance is read as logic 0. Two threshold resistance settings (TRSs) are marked as dashed lines in Fig. 5d. Fig. 5e illustrates the implementation of six universal Boolean logic functions in a single phase-transition spintronic device. Using AND as an example, AND is a logic function of two binary inputs. The output is always logic 0 except when the inputs are both logic 1, in which case the output is logic 1. TRS 1 is used in this case. An output resistance above TRS 1 is defined as logic 1, and a resistance below is logic 0; thus, resistance 1 (R1) is logic 1, and resistance 2 (R2), 3 (R3) and 4 (R4) are logic 0. When the illumination is off and a field is applied, logic 1 is the output. The detailed settings for this phase-transition device to realize different logic functions are presented in Fig. 5e. The arrow shows the resistance range in which the signal should be read as 1, and the opposite should be 0. Following the instructions given by the arrows in Fig. 5e, the logic functions of OR, NAND and NOR can be fulfilled. For XOR and NXOR, TRS 1 and TRS 2 are both used to define the resistance window. If the resistance between TRS 1 and TRS 2 is logic 1 and a resistance that exceeds this range is logic 0, then, R2 and R3 are logic 1, and R1 and R4 are logic 0. Thus, XOR is programmed. If the setting is reversed, NXOR is programmed. For NOT logic operation, the light input can be omitted. The logic operation can be achieved by changing the field from the $H_x$ to $H_y$ direction at the same field input setting (Fig. 5d). With the ability of responding to both optical and Oersted-field inputs, this optically controlled phase-transition spintronic device, which implements various logic functions in a single device, has great application potential for reconfigurable, programmable and cascadable logic devices beyond CMOS and will pave the way for big data and neuromorphic computing[37,38].

## DISCUSSION

In this work, we combine the strongly correlated electron material and the spintronic material to fabricate newly artificial heterostructures, which endow the spintronic material with optical modulation ability of magnetic and electrical properties by the reversible phase-transition process. The variation of the interfacial anisotropy energy, which reflects the modulation ability of the method, reaches 130 $\mu J/m^2$. It is comparable to the traditional modulation strategies based on strain[39,40] or charge mediated VCMA[3,41]. While this method possesses the potential advantages of optical modulation like ultralow-power and ultrahigh-speed. In the meantime, this optical control magnetism modulation strategy has little restriction on the choice of magnetic material, which indicates its advantage over other methods using selected materials (usually rare earth–transition metal (RE) based alloys, multilayers, heterostructures or special designed RE-free heterostructures[6,42]) to realize optical control. Furthermore, considering the phase transition of $VO_2$ can also be triggered by electrical gating, it is prospective to realize modulation of physical properties under the coordination of optical, electrical and magnetic fields in this phase-transition spintronic device, which is worthy of further research as a comprehensive topic.

To conclude, we have demonstrated optical control of electrical and magnetic modulation in a NiFe/$VO_2$ bilayer heterostructure. Phase-transition-induced changes in the electrical conductivity (55%), coercivity (60%) and magnetic anisotropy (33.5%) have been observed during the phase transition. Theoretical analyses and experimental evidence reveal that the magneto-elastic coupling induced by appreciable interfacial strain (up to 18.5%) may be the origin of the magnetism modulation. Further research on PTAMR devices fabricated with the heterostructure sample showed that multiresistance states and seven universal logic functions (AND, OR, NAND, NOR, XOR, NXOR and NOT) can be achieved in a single device. Our work, as a demonstration of optically controlled phase-transition spintronics, may pave the way for next-generation electronics.

## METHODS

### Heterostructure and device fabrication

The VO$_2$ thin films were grown on (001) and (100) TiO$_2$ substrates using a pulsed laser deposition (PLD) system (KrF, $\lambda$=248 nm). The growth of epitaxial VO$_2$ thin films was performed at 500 °C with 2.0 Pa of oxygen. The laser fluence and repetition rate were fixed at 4 J/cm$^2$ and 2 Hz, respectively. The film thickness was 20 nm after 7200 deposition pulses. The Ni$_{80}$Fe$_{20}$ films were magnetron-sputtered at room temperature. Raw bilayer samples were patterned by optical lithography (Micro Writer ML Baby, Durham Magneto Optics) followed by argon ion-beam etching into Hall bars with several legs with different spacings (Fig. 2a & 2b). Optical lithography and e-beam evaporation were then used to prepare Cr (5 nm)/Au (100 nm) electrodes. Then, 10 nm of SiO$_2$ was deposited by thermal evaporation on the device as a protection layer.

**Characterization and measurement**

The magnetic properties were measured by MOKE (NanoMOKE3, Durham Magneto optics ltd) with a red-light laser (LR-MFJ-660/2000 mW, Changchun Laser Technology Co.) heat source and SQUID (MPMS3, Quantum Design, Inc.) at room and high temperatures. The phase-transition AMR devices were measured by normal 4-terminal methods with Keithley 6221 and Keithley 2182 sourcing and measuring units, respectively. Meanwhile, an in-plane electromagnet (East Changing Co. China) provided magnetic fields with proper directions.

**REFERENCES AND NOTES**

1. Wolf, S. A. Spintronics: A Spin-Based Electronics Vision for the Future. *Science* **294,** 1488–1495 (2001).

2. Chappert, C., Fert, A. & Van Dau, F. N. The emergence of spin electronics in data storage. *Nat.*

Mater. **6,** 813–823 (2007).

3. Ong, P. V., Kioussis, N., Amiri, P. K. & Wang, K. L. Electric-field-driven magnetization switching and nonlinear magnetoelasticity in Au/FeCo/MgO heterostructures. *Sci. Rep.* **6,** 1–8 (2016).

4. Maruyama, T. *et al.* Large voltage-induced magnetic anisotropy change in a few atomic layers of iron. *Nat. Nanotechnol.* **4,** 158–161 (2009).

5. Stanciu, C. D. *et al.* All-optical magnetic recording with circularly polarized light. *Phys. Rev. Lett.* **99,** 1–4 (2007).

6. Mangin, S. *et al.* Engineered materials for all-optical helicity-dependent magnetic switching. *Nat. Mater.* **13,** 286–292 (2014).

7. Ramesh, R. & Spaldin, N. A. Multiferroics: Progress and prospects in thin films. *Nat. Mater.* **6,** 21–29 (2007).

8. Wu, S. M. *et al.* Reversible electric control of exchange bias in a multiferroic field-effect device. *Nat. Mater.* **9,** 756–761 (2010).

9. Wei, G. *et al.* Reversible control of magnetization of $Fe_3O_4$ by a solid-state film lithium battery. *Appl. Phys. Lett.* **110,** (2017).

10. Lu, N. *et al.* Electric-field control of tri-state phase transformation with a selective dual-ion switch. *Nature* **546,** 124–128 (2017).

11. Bauer, U. *et al.* Magneto-ionic control of interfacial magnetism. *Nat. Mater.* **14,** 174–181 (2015).

12. Liu, P. *et al.* Optically Tunable Magnetoresistance Effect: From Mechanism to Novel Device Application. *Materials (Basel).* **11,** 47 (2017).

13. Koch, T. L. & Koren, U. Semiconductor photonic integrated circuits. *IEEE J. Quantum Electron.*

**27,** 641–653 (1991).

14. Sun, X. *et al.* A molecular spin-photovoltaic device. *Science* **357,** 677–680 (2017).

15. Imada, M., Fujimori, A. & Tokura, Y. Metal-insulator transitions. *Rev. Mod. Phys.* **70,** 1039–1263 (1998).

16. Nakano, M. *et al.* Collective bulk carrier delocalization driven by electrostatic surface charge accumulation. *Nature* **487,** 459–462 (2012).

17. Lee, S. *et al.* Anomalously low electronic thermal conductivity in metallic vanadium dioxide. *Science* **355,** 371–374 (2017).

18. Jeong, J. *et al.* Suppression of Metal-Insulator Transition in $VO_2$ by Electric Field-Induced Oxygen Vacancy Formation. *Science* **339,** 1402–1405 (2013).

19. Tao, Z. *et al.* Decoupling of structural and electronic phase transitions in $VO_2$. *Phys. Rev. Lett.* **109,** 166406 (2012).

20. Cavalleri, A., Dekorsy, T., Chong, H. H. W., Kieffer, J. C. & Schoenlein, R. W. Evidence for a structurally-driven insulator-to-metal transition in $VO_2$: A view from the ultrafast timescale. *Phys. Rev. B - Condens. Matter Mater. Phys.* **70,** 1–4 (2004).

21. Kübler, C. *et al.* Coherent structural dynamics and electronic correlations during an ultrafast insulator-to-metal phase transition in $VO_2$. *Phys. Rev. Lett.* **99,** 1–4 (2007).

22. Liu, K., Lee, S., Yang, S., Delaire, O. & Junqiao, W. Recent progress on physics and applications of $VO_2$. *Mater. Today* in press (2018).

23. Wang, W. G., Li, M., Hageman, S. & Chien, C. L. Electric-field-assisted switching in magnetic tunneljunctions. *Nat. Mater.* **11,** 64–68 (2012).

24. Ikeda, S. *et al.* A perpendicular-anisotropy CoFeB–MgO magnetic tunnel junction. *Nat. Mater.*

**9,** 721–724 (2010).

25. Engel, B. N., England, C. D., Van Leeuwen, R. A., Wiedmann, M. H. & Falco, C. M. Interface magnetic anisotropy in epitaxial superlattices. *Phys. Rev. Lett.* **67,** 1910–1913 (1991).

26. Song, O., Ballentine, C. A. & O'Handley, R. C. Giant surface magnetostriction in polycrystalline Ni and NiFe films. *Appl. Phys. Lett.* **64,** 2593–2595 (1994).

27. Reekstin, J. P. Zero Magnetostriction Composition of NiFe Films. *J. Appl. Phys.* **38,** 1449–1450 (1967).

28. Tretiakov, O. A., Morini, M., Vasylkevych, S. & Slastikov, V. Engineering Curvature-Induced Anisotropy in Thin Ferromagnetic Films. *Phys. Rev. Lett.* **119,** 077203 (2017).

29. Graczyk, P., Schäfer, R. & Mroz, B. Magnetoelastic coupling between NiFe thin film and $LiCsSO_4$ studied by Kerr microscopy. *J. Phys. D. Appl. Phys.* **48,** 425002 (2015).

30. Garshelis, I. J. Force transducers based on the stress dependence of coercive force. *J. Appl. Phys.* **73,** 5629–5631 (1993).

31. Garrett, C. *et al.* Thickness dependence of the magnetic hysteresis of NiFe-31% films as a function of an applied isotropic in-plane stress. *J. Appl. Phys.* **93,** 8624–8626 (2003).

32. Ma, H. *et al.* Flexible, All-Inorganic Actuators Based on Vanadium Dioxide and Carbon Nanotube Bimorphs. *Nano Lett.* **17,** 421–428 (2017).

33. Xiao, L. *et al.* Fast Adaptive Thermal Camouflage Based on Flexible $VO_2$/Graphene/CNT Thin Films. *Nano Lett.* **15,** 8365–8370 (2015).

34. Dong, K. *et al.* A Lithography-Free and Field-Programmable Photonic Metacanvas. *Adv. Mater.* **30,** 1703878 (2018).

35. Lin, X. *et al.* Gate-Driven Pure Spin Current in Graphene. *Phys. Rev. Appl.* **8,** 034006 (2017).

36. Wan, C. *et al.* Programmable Spin Logic Based on Spin Hall Effect in a Single Device. *Adv. Electron. Mater.* **3,** 1600282 (2017).

37. Xia, Q. *et al.* Memristor - CMOS Hybrid Integrated Circuits for Reconfigurable Logic. *Nano Lett.* **9,** 3640–3645 (2009).

38. Coskun, A., Deniz, E. & Akkaya, E. U. Effective PET and ICT switching of boradiazaindacene emission: A unimolecular, emission-mode, molecular half-subtractor with reconfigurable logic gates. *Org. Lett.* **7,** 5187–5189 (2005).

39. Yu, G. *et al.* Strain-induced modulation of perpendicular magnetic anisotropy in Ta/CoFeB/MgO structures investigated by ferromagnetic resonance. *Appl. Phys. Lett.* **106,** (2015).

40. Mardana, A., Ducharme, S. & Adenwalla, S. Ferroelectric control of magnetic anisotropy. *Nano Lett* **11,** 3862–7 (2011).

41. Hibino, Y. *et al.* Peculiar temperature dependence of electric-field effect on magnetic anisotropy in Co/Pd/MgO system. *Appl. Phys. Lett.* **109,** 1–5 (2016).

42. Becker, J. *et al.* Ultrafast Magnetism of a Ferrimagnet across the Spin-Flop Transition in High Magnetic Fields. **117203,** 1–5 (2017).

## Acknowledgments

The authors thank Huaiwen Yang, Jiwei Hou, Runrun Hao, and Junfeng Qiao for their help in device fabrication. This work was supported by the National Natural Science Foundation of China (Nos. 51602013, 61704005, 61571023 and 61627813), the International Collaboration 111 Project (No. B16001), the Beijing Natural Science Foundation (No. 4162039), the China Postdoctoral Science Foundation (No.

**Author contributions**

G.D.W. and X.Y.L. contributed equally to this work. X.Y.L. and W.S.Z. coordinated the project. X.Y.L. proposed and designed the research. G.D.W., Z.Z.S., D.W., X.H.W., K.L.J., K.L. and Y.X.C. performed the sample preparation. G.D.W., Z.Z.S. and X.Y.L performed the heterostructure characterization and measurement. G.D.W., Z.Z.S. and X.Y.L. performed the device fabrication and measurement. G.D.W, X.Y.L., Y.X.C. and W.S.Z. contributed to the mechanism interpretation. G.D.W., X.Y.L. and W.S.Z. wrote the paper. All the authors participated in discussions on the research.

**Additional information**

**Supplementary Information** accompanies this paper at

http://www.nature.com/naturecommunications

**COMPETING FINANCIAL INTERESTS STATEMENT**

The authors declare no competing financial interests.

# Figures and Tables

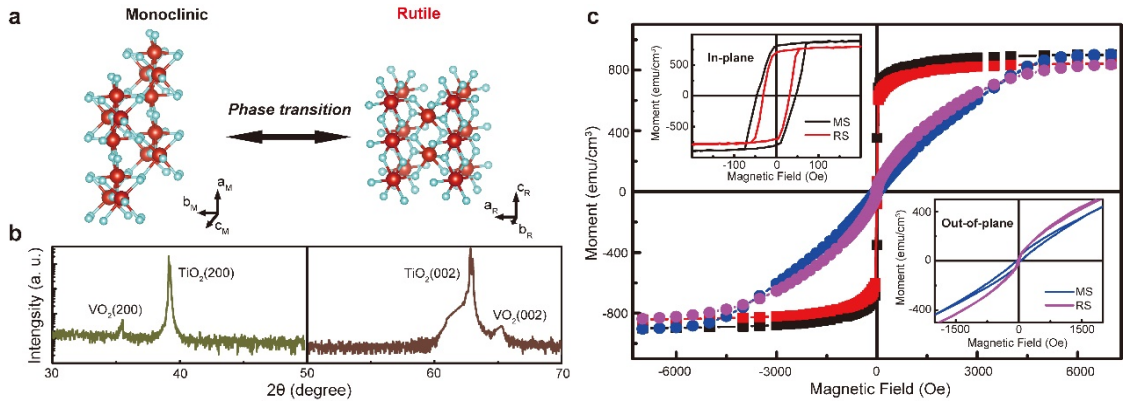

**Figure 1 | Phase transition of VO$_2$ and characterization of the NiFe/VO$_2$ heterostructure.** (**a**) Schematic illustration of the reversible phase transition of VO$_2$ between monoclinic (M1) and rutile (R) lattice structures. (**b**) XRD *θ-2θ* scan of the NiFe/VO$_2$ heterostructure grown on TiO$_2$ (100) and TiO$_2$ (001) substrates, suggesting epitaxial growth of the VO$_2$ layer and amorphous growth of the NiFe layer. (**c**) *M–H* hysteresis loops of the NiFe (5 nm)/VO$_2$ (20 nm)/TiO$_2$ (100) sample measured before and after the phase transition of VO$_2$ (MS at 300 K and RS at 360 K). The magnetic field was applied perpendicular (data plotted as circles) and parallel (data plotted as squares) to the surface. The insets give the scanning results of the small fields.

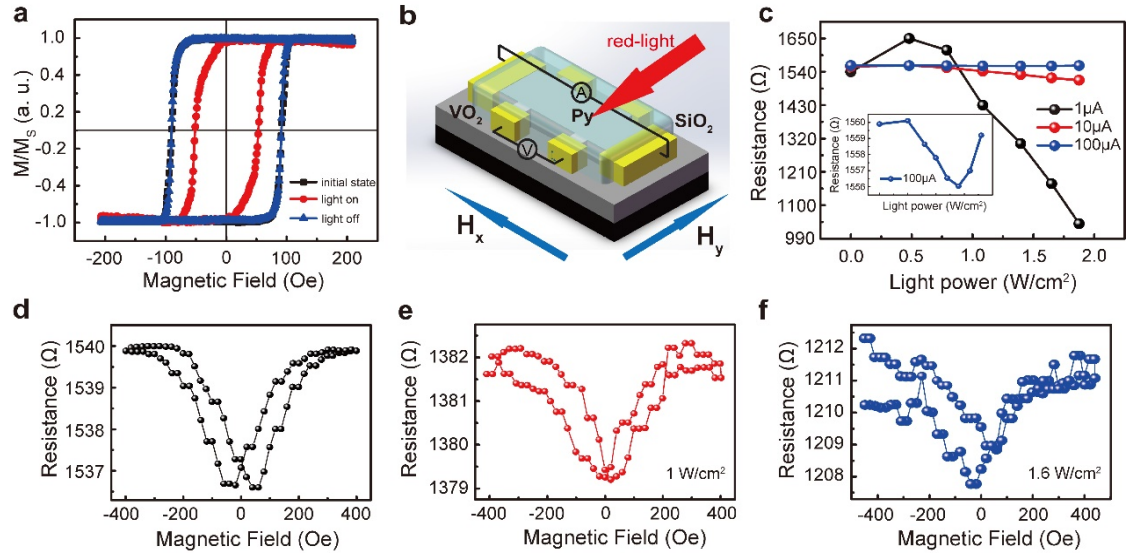

**Figure 2 | Optically controlled magnetoresistance measurements.** (**a**) The hysteresis loops of the sample at initial, light-on and light-off states. (**b**) Schematic drawing of the PTAMR device. Direction of the applied magnetic fields, $H_x$ and $H_y$, are also present. (**c**) The device resistance change measured with varying applied current under increasing illumination power. (**d, e &f**) Comparison of the $H_x$ magnetoresistance of the NiFe (5 nm)/VO$_2$ (20 nm)/TiO$_2$ (100) device without (d) and with the illumination of 1-W/cm$^2$ (e) and 1.6-W/cm$^2$ (f) red light.

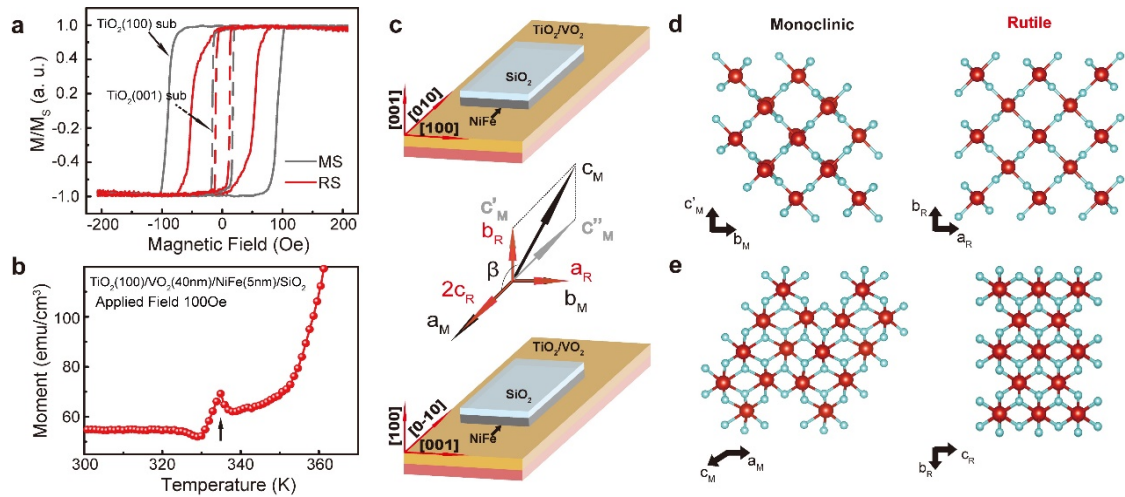

**Figure 3 | Interface strain analysis and characterization of magnetism modulation.** (**a**) Longitudinal Kerr signals of the NiFe/VO$_2$ heterostructure grown on different substrates. The $H_C$ change between the MS and RS is 45% for the TiO$_2$ (100) substrate and 25% for the TiO$_2$ (001) substrate when the illumination power reaches 2 W/cm$^2$. (**b**) Temperature dependence of out-of-plane magnetization with a magnetic field of 100 Oe. The arrow marks the VO$_2$ phase-transition temperature. (**c**) Strain analysis of the interface between VO$_2$ and NiFe when the VO$_2$ phase transition occurs. (**d & e**) Schematic diagram of the crystal lattice change of the (001) plane and (100) plane with regard to the rutile structure.

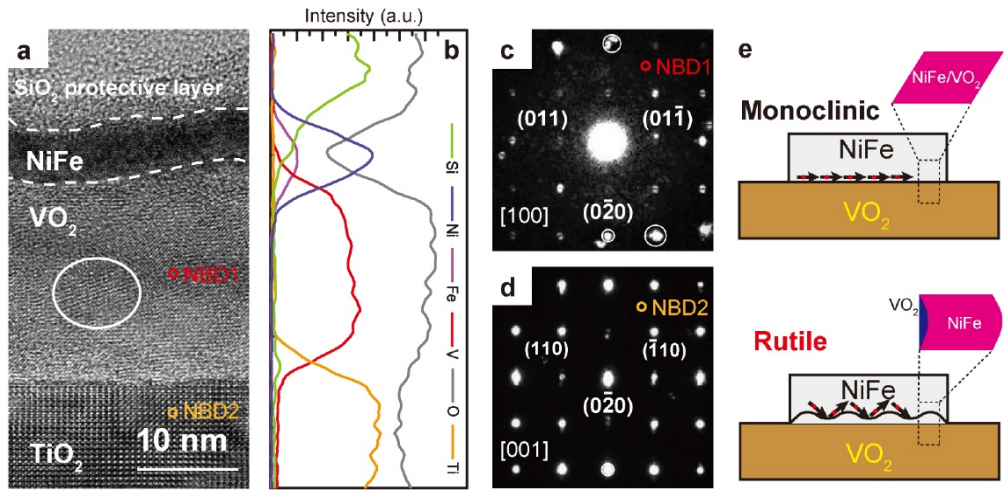

**Figure 4 | High-resolution transmission electron microscopy (HR-TEM) analysis of the NiFe/VO$_2$ heterostructure.** (**a**) Cross-section HR-TEM image of the NiFe/VO$_2$ heterostructure deposited on a TiO$_2$ (100) substrate. Some distortion of the crystal orientation can be detected in the VO$_2$ layer, as marked in this figure. (**b**) Energy dispersive spectroscopy (EDS) analysis of different elements (O, Ti, Fe, Si, V, and Ni) in the heterostructure. (**c & d**) Nanobeam electron diffraction (NBD) corresponding to NBD1 and NBD2 in (b). (**e**) Schematic diagram of the heterostructure. When the crystal lattice changes, the increased interface roughness causes the spin, which was originally pinned in-plane, to turn to the out-of-plane direction.

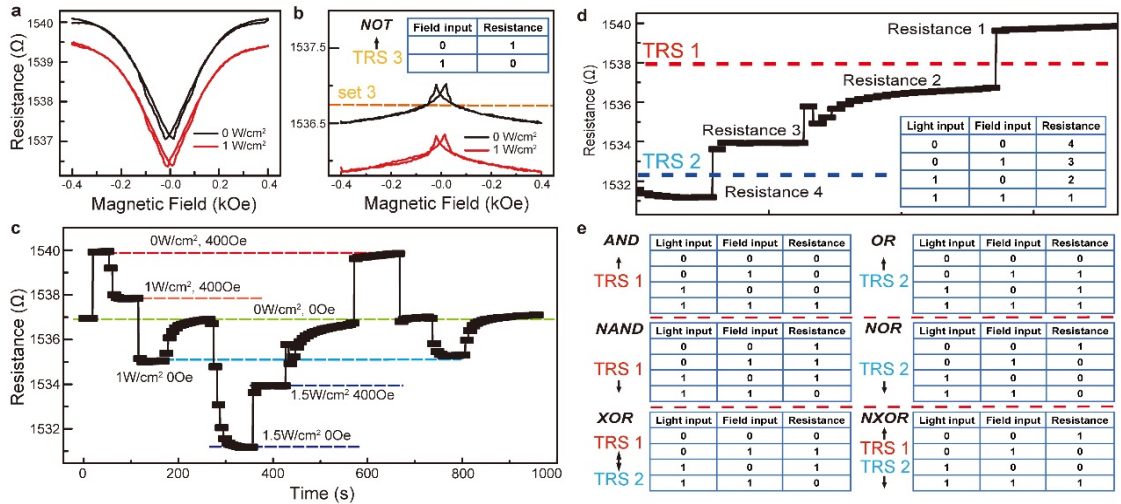

**Figure 5 | Optically controlled multiresistance states and programmable logic implementation.** (**a & b**) Magnetoresistance curves measured with 0-W/cm$^2$ and 1-W/cm$^2$ light illumination using 10 μA as the measuring current. The magnetic field is applied in the $H_x$ (a) and $H_y$ (b) directions. (**c**) Six different resistance states are realized using two kinds of illumination power and $H_x$=400 Oe magnetic field. (**d**) Four resistance states are selected from (c) to illustrate the logic operation. Light illumination and magnetic fields are designed as two kinds of input signals. Illumination off (0 W/cm$^2$) is set as logic 1 and on (1 W/cm$^2$) as logic 0, while the field applied (400 Oe) is set as logic 1 and no field (0 Oe) as logic 0. The result is read out by the resistance and shown in the truth table. Two threshold resistance settings (TRSs) are marked as red and blue dashed lines. (**e**) The truth table of six basic logic operations with different TRSs. The arrow shows the resistance range for which the signal should be 1, and the opposite should be 0. NOT operation can be achieved by turning the field to $H_y$ with the same setting as the field input. The threshold voltage and truth table are shown alone in (b).

Table 1. Magnetic properties of SiO$_2$ (20 nm)/NiFe (5 nm)/VO$_2$ (20 nm)/TiO$_2$ (100)

| Temperature | $H_C$ | $M_S$ | $H_K$ [a] | $K_u$ | $K_i$ |
|---|---|---|---|---|---|
| K | Oe | emu/cm$^3$ | Oe | erg/cm$^3$ | erg/cm$^2$ |
| 300 | 72 | 898 | 4365 | -3.55×10$^6$ | 0.755 |
| 360 | 45 | 840 | 4229 | -2.66×10$^6$ | 0.885 |

[a] $H_K$ was obtained by extracting the field corresponding to 90% of the out-of-plane moment.

# Supplementary Materials

## Section Ⅰ. Characterization of VO$_2$

Temperature control and red-light illumination are used as different stimuli for the phase transition of VO$_2$. As illustrated in Fig. S1, the resistance variation of VO$_2$ grown on a TiO$_2$ (100) substrate reached three orders of magnitude. When temperature is used to trigger the transition (Fig. S1a), the critical temperature is near 340 K, which is consistent with the magnetic change of the heterostructure described in Fig. 4e in the main text. For the illumination-triggered phase transition (Fig. S1b), the most intense resistance change occurs from 500 to 750 mW, and an obvious variation loop could be detected, as observed for the thermal methods. The temperature dependence of the out-of-plane magnetization at a magnetic field of 100 Oe was also measured, and no magnetic phase was found in this VO$_2$ sample.

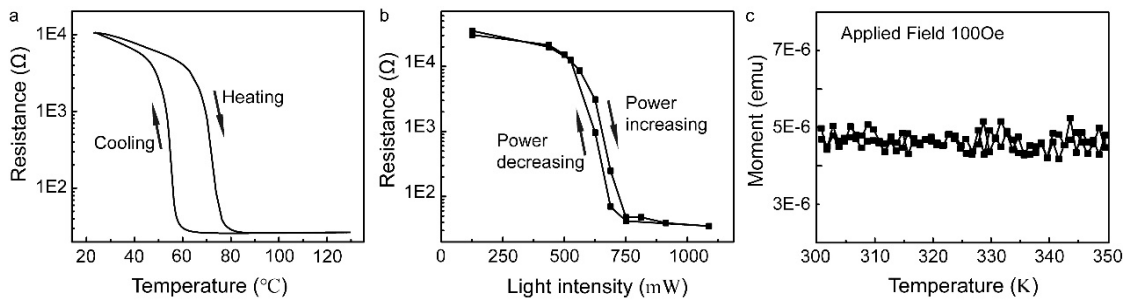

**Figure S1 | Characterization of VO$_2$ grown on a TiO$_2$ (100) substrate.** (**a**) The phase transition triggered by temperature. The critical temperature is near 70 ℃ (343 K). (**b**) The phase transition triggered by red-light illumination. The most intense phase transition occurs from 500 to 750 mW, but the resistance still changes beyond this range. (**c**) Temperature dependence of the out-of-plane magnetization at a magnetic field of 100 Oe. The signal reached the detectable limit of the instrument, and no magnetic phase was found.

## Section II. Coercivity change of different samples

*In situ* longitudinal magneto-optic Kerr effect (MOKE) measurements using red-light illumination to trigger the phase transition were performed for several samples. Fig. S2a shows results for the contrast NiFe (5 nm) sample grown on a $SiO_2$ substrate. No obvious $H_C$ change can be observed, apart from faint deviations caused by the thermal effect. Fig. S2b presents the hysteresis curves of NiFe (3 nm)/$VO_2$ (40 nm)/$TiO_2$ (100). It is found that the change rate of $H_C$ reaches 67% with the light on and off, which is slightly higher than that of the 5-nm sample in the main text. We changed the thickness of the NiFe layer to 10 nm (Fig. S2c), and the change rate quickly dropped to 0%. This thickness dependence indicates that the interfacial strain coupling plays a crucial role in the magnetism modulation.

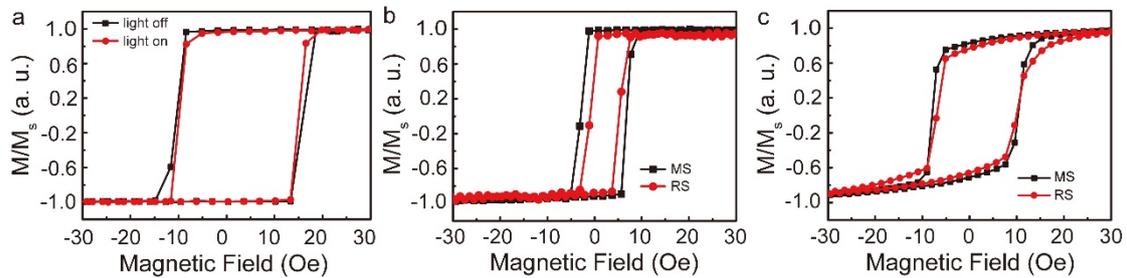

**Figure S2 | MOKE measurements using red-light illumination to trigger the phase transition.** (**a**) MOKE measurement results of NiFe (5 nm)/$SiO_2$. (**b**) MOKE measurement results of NiFe (3 nm)/$VO_2$ (40 nm)/$TiO_2$ (100). The change rate of $H_C$ reaches 67% between the on and off conditions. (**c**) MOKE measurement results of NiFe (10 nm)/$VO_2$ (40 nm)/$TiO_2$ (100).

**Section III. Lattice parameter change calculation**

The structure variation of $VO_2$ during the phase transition is qualitatively analyzed based on the bulk parameters. As the temperature increases, the monoclinic structure of $VO_2$ transforms to the rutile structure, and vice versa for a temperature decrease. The change rates in each direction can be calculated as $k = \frac{l_M - l_R}{l_R}$, in which $l_M$ is the lattice parameter of the monoclinic structure and $l_R$ is the lattice parameter of the rutile structure. The calculated results are summarized in Table S1. As the transition occurs, $a_M$ and $b_M$ vary along their original direction, crossing the structure variation and coinciding with the direction of $c_R$ and $a_R$. Thus, the change rates are 0.8% and -0.4%. As the angle $\beta$ decreases from 122.6° to 90°, the $c_M$ axis turns to the direction of $b_R$. To calculate the change rates along each direction, we decompose $c_M$ into $c_M'$ and $c_M''$ in the direction of the $b_R$ and $a_R$ axes. In the $TiO_2$ (001) plane, the spacing between atoms along $b_R$ is shortened from $c_M'$ to $b_R$, and the change rate is 0.2%. Moreover, there is a stress perpendicular to the plane due to the change in $c_M''$, causing the original uneven surface to become flat. In the $TiO_2$ (100) plane, the spacing between atoms along $b_R$ changes from $c_M$ to $b_R$, reaching 18.5%. In addition, there should be a shear force caused by the rotation of $c_M$.

**Supplementary TABLE S1. $VO_2$ lattice parameter change during the phase transition**

| Lattice axis<br>Crystal structure | $a_M/2c_R$<br>nm | $b_M/a_R$<br>nm | $c_M/b_R$<br>nm | $\beta$<br>° | $c_M'/b_R$<br>nm | $c_M''$<br>nm |
|---|---|---|---|---|---|---|
| Monoclinic | 0.575 | 0.452 | 0.538 | 122.6 | 0.453 | 0.290 |
| Rutile | 0.570 | 0.454 | 0.454 | 90 | 0.454 | - |